# *Physical essence of propagable fractional-strength optical vortices in free space*

Xiaoyu Weng[1]*, Yu Miao[2], Yang Li[2], Xiangmei Dong[2], Xiumin Gao[2]* and Songlin Zhuang[2]

*1 Shenzhen Key Laboratory of Photonics and Biophotonics, Key Laboratory of Optoelectronic Devices and Systems of Ministry of Education and Guangdong Province, College of Physics and Optoelectronic Engineering, Shenzhen University, Shenzhen, 518060, China*

*2 Engineering Research Center of Optical Instrument and System, Ministry of Education, Shanghai Key Lab of Modern Optical System, School of Optical-Electrical and Computer Engineering, University of Shanghai for Science and Technology, 516 Jungong Road, Shanghai 200093, China.*

\* Correspondence and requests for materials should be addressed to X.W. (email: xiaoyu@szu.edu.cn) or to X.G. (email: gxm@usst.edu.cn).

## Abstract:

Fractional-order vector vortex beams are recently demonstrated to be new carriers of fractional-strength optical vortices. However, why can those new vortex beams formed by the combination of both unstable states propagate stably in free space? Here, we solve this scientific problem by revealing the physical essence of propagable fractional-strength optical vortices in free space. Three new understandings regarding those peculiar vortex beams are therefore proposed, namely Abbe's diffraction limit, phase evolution of vortex beam, and phase's binary time vector property. For the first one, owing to Abbe's diffraction limit, the inherent polarization modes are intertwined together, thereby maintaining the entire peculiar vortex beams in free space. For the second one, we demonstrate the phase evolution of vortex beam, which is the physical reason of polarization rotation of fractional-order VVBs. For the third one, the phase is not merely a scalar attribute of light beam, but manifests a binary time vector property. This work provides entirely different physical viewpoints on the phase of vortex beam and Abbe's diffraction limit, which may deepen our knowledge on the behavior of light beam in classical optics.

# Introduction

A vortex beam refer to a light beam carrying optical vortex exp($il\varphi$) [1]. Here, $\varphi$ and $l$ are the vortex phase and its topological charge, respectively. Generally, conventional vortex beams, e.g., the Bessel-Gaussian beam and Laguerre-Gaussian beam, are special solutions of wave function, which always possess integer topological charge $l$ [2]. Therefore, optical vortices carrying by those light beams can not only induce orbital angular momentum (AM) equivalent to $l\hbar$ per photon, but also maintain stable in free space during propagating in free space [1]. Why is it so important that whether an optical vortex can be propagable in free space? The main reason is that information adhered to propagable vortex beams can transfer without distortion in free space [3]. For over 30 years, great success has been achieved in the study of optical vortices, including spin and orbital AM conversion [4, 5], rotational Doppler effects [6, 7], advanced lasers [8], optical imaging [9], and orbital AM optical communication [3, 10, 11]. However, most studies regarding orbital AM are still conducted under the framework of vortex beams with integer orbital AM.

One may wonder whether there is another kind of propagable optical vortices in optics that can not only possesses a noninteger orbital AM, but also propagate in free space stably like the above integer one. Recently, we solve this scientific problem by demonstrating propagable vortex beam with $l$+0.5 topological charge [12, 13]. It should be emphasized that this kind of vortex beam is also the solution of wave function. That is, optical vortices with fractional topological charge are naturally existed in free space. Generally, conventional vortex beam with integer orbital AM always possesses a homogeneous polarization state, such as linear, circular and elliptic polarization [1, 2]. Unlike the conventional one, fractional optical vortices have different carriers, namely VVBs with $m$+0.5 order [12]. Although our previous work provides a mathematical derivation on the creation of propagable fractional optical vortices, the physical essence of those peculiar vector beams is unclear. We don't know why the combination of both unstable states, namely $m$+0.5-order VVB and vortex phase with $l$+0.5 topological charge, can form a stable vortex beam with fractional-order and fractional topological charge. We don't know why the polarization and vortex phase of light beam remain stable in free space, but the discontinuity of light beam caused by the phase and polarization jump are missing. We even don't know why the modulation symmetry of light beam is broken, however, the inherent polarization modes within fractional-order VVB are still intertwined with each other [14].

Here, we reveal the physical essence of propagable vortex beam with $l$+0.5 topological charge by re-understanding Abbe's diffraction limit, investigating the phase evolution of vortex beam and proposing phase's binary time vector property. For this first one, although the modulation symmetry of fractional-order VVB is naturally broken down by the optical vortices with fractional topological charge, the inherent polarization modes are still intertwined with each other because of Abbe's diffraction limit, thereby leading to a stable state of fractional-order VVBs in free space. For the second one, we demonstrate the phase evolution of vortex beam, which is the physical reason of polarization rotation of fractional-order VVBs. For the third one, the phase is not merely a scalar attribute of light beam, but manifests a binary time vector property. Therefore, the entire electric field of fractional-order VVBs is continuous while the topological charge of vortex phase and the order of polarization are fraction. This work provides entirely different physical viewpoints on the propagable vortex beam with $l+0.5$ topological charge, which may deepen our knowledge on the behavior of light beam in classical optics.

## 2. Results

As two stable states in classical optics, vortex beam with integer topological charge and integer-order VVB represent two extremes of light beam. For the former one, vortex beam with integer topological charge is normally homogenously polarized, which can be considered as a scalar light beam. Therefore, those light beams can carry a convectional optical vortex with integer topological charge stably in free space. For the latter one, integer-order VVB can be obtained by the combination of left and right circularly polarized beams with inverse optical vortices. That is, the inherent modulation symmetry of light beam can be broken down by imposing an additional optical vortex [14]. For this reason, integer-order VVB can only maintain stable without optical vortex in free space. Unlike the both light beams, propagable vortex beam with $l$+0.5 topological charge can be considered as their middle state. The stability of these peculiar vortex beams is dependent on the interaction between the vortex phase and the polarization of fractional-order VVB. Without one of both, fractional-order VVB and vortex beam with non-integer topological charge cannot propagate individually in free space. However, why does this middle state of light beam exist naturally in free space?

### 2.1 Intertwinement of inherent polarization modes

Mathematically, propagable vortex beam with natural non-integer topological charge can be expressed as [12]

$$\mathbf{E} = \exp[i(l+0.5)\varphi]\begin{bmatrix} \cos[(m+0.5)\varphi+\beta] \\ \sin[(m+0.5)\varphi+\beta] \end{bmatrix}, \tag{1}$$

where $\varphi$ is the azimuthal angle; $l$ and $m$ are two integers that relates to the topological charge and order of entire vortex beam. $\beta$ determines the polarization direction of $m$+0.5-order VVB. Without loss of generality, $\beta = 0$ in this paper. Generally, the above vortex beam can be simplified into

$$\mathbf{E} = \exp[i(m+l+1)\varphi]|\mathbf{R}\rangle + \exp[-i(m-l)\varphi]|\mathbf{L}\rangle, \tag{2}$$

Here, $|\mathbf{L}\rangle = \begin{bmatrix} 1 & i \end{bmatrix}^t$ and $|\mathbf{R}\rangle = \begin{bmatrix} 1 & -i \end{bmatrix}^t$ denote the left and right circularly polarized modes, respectively. $t$ denotes the matrix transpose operator.

As demonstrated in our previous work, Equation (2) represents stable propagable vortex beams with fractional topological charge [12]. Therefore, the inherent polarization modes in Eq. (2) are always intertwined together even when the light beam propagate to an infinite distance in free space. However, in Eq. (2), the topological charge of $|\mathbf{L}\rangle$ are different from that of $|\mathbf{R}\rangle$, different topological charge leads to different position of $|R\rangle$ and $|L\rangle$, and the maximum distance between both polarized modes can be obtained at the infinite distance in free space. That is, the first scientific problem is that how can both circular polarization modes in Eq. (2) intertwine together so that the entire vortex beam remains stable in free space.

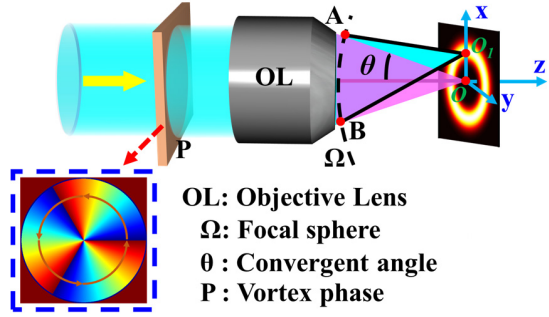

**Figure 1.** Schematic of vortex beam with different topological charge in focusing system. $\Omega$ is the focal sphere of objective lens (OL), with its center at O and a radius $f$. A, B are two arbitrary points in $\Omega$. $O_1$ is the point on vortex beam in the focal plane. P is the vortex phase of incident light beam. $\theta$ is the convergent angle of OL.

To verify the intertwinement of inherent polarization modes in Eq. (2), we explore the effect of vortex phase affecting $|L\rangle$ and $|R\rangle$ when the entire vortex beam propagates to an infinite distance in free space. According to information optics, the image of a light beam at an infinite distance can simply be obtained in the focal region of an objective lens (OL) obeying the sine condition in Fig. 1 [15]. Here, $\Omega$ is the focal sphere, and its center is at O and its radius is $f$, which is the focal length of OL. $O_1$ is an arbitrary point in the focal region of the lens. As shown in Fig. 1, when the topological charge of vortex phase (P) is zero, constructive interference can only occur at the focal point O owing to the equivalent optical paths of the light beams between the points in $\Omega$ and point O. Therefore, a bright focal spot can be obtained at point O. By contrast, the light intensity of focus became a donut when the topological charge is an integer $l$. It should be emphasized that this donut in the focal plane indicates the image of vortex beam propagating to an infinite distance in free space.

Due to the phase symmetry of vortex beam, the maximum light intensities on the donut are located at the point $O_1$ (see Fig. 1), the position of which are merely determined by the optical path difference (OPD) for the light beams between two arbitrary symmetry points A, B in $\Omega$ and $O_1$. Note that light beams of other points in $\Omega$ are interfered at the donut with different weight factor according to the principle of interferometry. In the cylindrical coordinate system, points A, B, O and $O_1$ can be expressed as $(f\sin\theta\cos\varphi_A, f\sin\theta\sin\varphi_A, -f\cos\theta)$, $(f\sin\theta\cos\varphi_B, f\sin\theta\sin\varphi_B, -f\cos\theta)$, $(0,0,0)$ and $(\rho\cos\varphi_s, \rho\sin\varphi_s, z)$, respectively. Here, $\varphi_A = \varphi_B - \pi = \varphi$, and $z=0$ indicates that the vortex beam is in the focal plane.

The OPD for the light beams between the points A, B in Ω and $O_1$ can be simplified to $L_2-L_1$, where $L_1$ and $L_2$ denote the light paths $AO_1$ and $BO_1$, respectively. Generally, the light paths $L_1$ and $L_2$ can be expressed as

$$L_1 = \sqrt{(f\sin\theta\cos\varphi - \rho\cos\varphi_s)^2 + (f\sin\theta\sin\varphi - \rho\sin\varphi_s)^2 + (f\cos\theta)^2} \quad (3)$$

$$L_2 = \sqrt{(-f\sin\theta\cos\varphi - \rho\cos\varphi_s)^2 + (-f\sin\theta\sin\varphi - \rho\sin\varphi_s)^2 + (f\cos\theta)^2} \quad . \quad (4)$$

The OPD for the light beams between the points in Ω and $O_1$ can be calculated as

$$\Delta s = L_2 - L_1 = \frac{4\rho\sin\theta\cos(\varphi - \varphi_s)}{\sqrt{1 + \eta_\rho^2 + 2\eta_\rho\sin\theta\cos(\varphi - \varphi_s)} + \sqrt{1 + \eta_\rho^2 - 2\eta_\rho\sin\theta\cos(\varphi - \varphi_s)}} \quad (5)$$

where $\theta$ is the convergent angle; $\eta_\rho = \rho/f$. Because $O_1$ is the point on the vortex beam in the vicinity of the focus, $\rho \ll f$. That is, $\eta_\rho \approx 0$. Finally, Equation 5 can be simplified to

$$\Delta s = 2\rho\sin\theta\cos(\varphi - \varphi_s). \quad (6)$$

It should be emphasized that $\varphi$ and $\varphi_s$ are the azimuthal angles in the wavefront and in the focal plane of OL, respectively. What is the relationship between both of them?

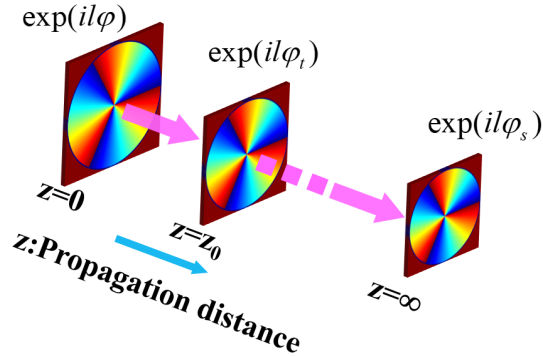

**Figure 2.** Propagation invariance of optical vortices with integer topological charge. During propagating in free space, vortex phase of light beam with integer topological charge maintain stably. Here, $\varphi$, $\varphi_t$ and $\varphi_s$ are the vortex phases in different propagation distance.

In principle, vortex beams with integer topological charge are special solutions of wave function. As shown in Fig. 2, optical vortex carried by a light beam can maintain itself in free space even when it propagates to an infinite distance. For this reason, one can obtain the same vortex phase at arbitrary propagation distance in free space. In the case of $l \geq 0$, the rotation direction of $\varphi$ and $\varphi_s$ are homodromous, therefore $\varphi = \varphi_s$. In the case of $l < 0$, the rotation direction of $\varphi$ and $\varphi_s$ are inverse. Due to the symmetric of vortex phase, $\varphi = \varphi_s + l\pi$.

Finally, Equation 6 can be expressed as

$$\Delta s = \begin{cases} 2\rho \sin\theta & l \geq 0 \\ 2\rho \sin\theta \cos(l\pi) & l < 0 \end{cases}. \tag{7}$$

According to the principle of interferometry, when modulating by a vortex phase (P) $\exp(il\varphi)$ in Fig. 1, the above OPD for the light beams in Eq. (7) are compensated by the vortex phase $\exp(il\varphi)$, which can be expressed as

$$\Delta s - l\lambda/2 = M\lambda. \tag{8}$$

Here, $l$ is the topological charge of vortex beam; $M$ indicates the order of constructive and deconstructive interference on vortex beam. Further, Equation 8 can be simplified into

$$\Delta s = |M + l/2|\lambda. \tag{9}$$

It should be emphasized that high order constructive interference of vortex beam always occurs at the outer ring. For this reason, for $l > 0$, $M > 0$ and for $l < 0$, M < 0.

From Eqs. (7) and (9), one can simply obtain the radius $\rho_M$ of maximum light intensities of vortex beam in the focal plane of OL, namely OO$_1$ in Fig. 1, which can be expressed as

$$\rho_{M,l} = \frac{|M + l/2|\lambda}{2\sin\theta}. \tag{10}$$

For more convenient, $\rho_{M,l}$ and $\rho_{M+0.5,l}$ denotes the radius of bright rings (constructive interference) and dark rings (deconstructive interference) of vortex beam with topological charge $l$, respectively. Here, $M = 0, \pm 1, \pm 2...$. Note that $|\cos(l\pi)| = 1$ with $l<0$.

Equation 10 not only implies that vortex beam with topological charge $l$ are composed of a series of bright and dark rings, but also demonstrates that vortex beams with topological charge $\pm l$ possess an identical radius $\rho_{M,\pm l}$. Therefore, left and right circular polarization mode with inverse optical vortices are always intertwined together, thereby generating a stable conventional integer-order VVB in free space [16]. Here, the order is determined by the topological charge of both circular polarized beams. In term of vortex beam with fractional topological charger in Eq. (2), the inherent left and right circular polarization mode possess different topological charge, thereby leading to different trajectory of bright rings in the focal plane of OL. Without loss of generality, we take the brightest ring as example to explain how both circular intertwine together to form a stable vortex beam in Eq. (2).

Figure 3 presents the brightest ring with radius $\rho$ of vortex beam with $M=0$ in Eq. (10). According to Eq. (10), $\rho_{M,l} = \dfrac{|M+l/2|\lambda}{2\sin\theta}$ denotes the radius of bright ring and $\rho_{M+0.5,l=0} = \dfrac{|M+0.5+l/2|\lambda}{2\sin\theta}$ represents the radius of dark ring. Finally, the thickness of each bright ring of vortex beam can simply be obtained by

$$D = 2(\rho_{M,l} - \rho_{M+0.5,l}) = \frac{\lambda}{2\sin\theta} \tag{11}$$

In the condition of $l=0$, a bright spot is obtained in the focal plane. That is, $\rho_{M=0,l=0} = 0$ and $\rho_{M=0.5,l=0} = \dfrac{\lambda}{4\sin\theta}$ for dark ring of vortex beam. That is, one can derive the Abbe's diffraction limit in free space by calculating the size of focus in Fig. 3(a) using Eq. (11), namely $2\rho_{M=0.5,l=0} = \dfrac{\lambda}{2\sin\theta}$.

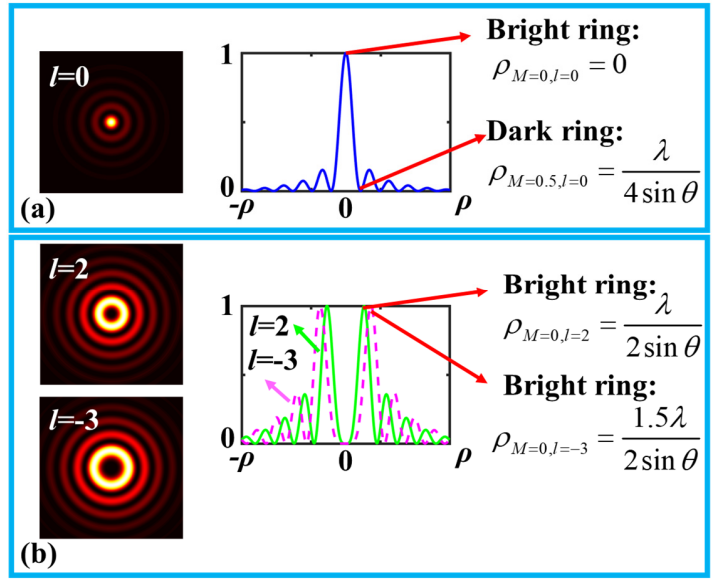

**Figure 3.** Schematic of vortex beam with different radius $\rho$. (a) denotes the light beam with topological charge $l=0$; (b) represents two adjacent vortex beams with $l=2$ and $l=-3$. The combination of both vortex beam can form an m=-2.5-order VVB with topological charge $l=-0.5$ in Eq. (2).

In the condition of $l \neq 0$, the vortex beam becomes a donut in the focal plane. From Eq. (11), all bright rings of vortex beam share an identical thickness, which is consistent with the Abbe's diffraction limit in free space, as shown in Fig. 3 (b). Although the brightest rings with a different of l possesses different trajectory in the focal plane, these two adjacent brightest rings cannot distinguish from each other because of Abbe's diffraction limit, as shown in Fig. 3(b). That is why the modulation symmetry of light beam is broken, however, the inherent left and right polarization modes with the topological charge difference of 1 are still intertwined together. For this reason, Equation 1 merely presents two stable vortex beams in free space. One is a vortex beam that possesses an optical vortex with topological charge ±0.5 along with a polarization state of an *m*+0.5-order VVB (see Equation (12)); another is a vortex beam that has a constant polarization state, namely, a ±0.5-order VVB, but an optical vortex with arbitrary topological charge *l*+0.5 (see Equation (13)) [12, 13].

$$\mathbf{E}_{m+0.5} = \exp(\pm i 0.5\varphi) \begin{bmatrix} \cos[(m+0.5)\varphi + \beta] \\ \sin[(m+0.5)\varphi + \beta] \end{bmatrix}, \quad (12)$$

$$\mathbf{E}_{l+0.5} = \exp[i(l+0.5)\varphi] \begin{bmatrix} \cos(\pm 0.5\varphi + \beta) \\ \sin(\pm 0.5\varphi + \beta) \end{bmatrix}, \quad (13)$$

For many years, we focus on how to break Abbe's diffraction limit so that high resolution can be achieved in optical imaging, optical lithography. However, in term of light beam generation, Abbe's diffraction limit is the physical essence of the intertwinement of inherent left and right circular polarization modes in free space. That is, the existence of this propagable vortex beam with *l*+0.5 topological charge benefits from the Abbe's diffraction limit. Without Abbe's diffraction limit, the intertwinement inherent polarization modes cannot occur, and these peculiar vortex beams cannot be created in free space.

## 2.2 Phase of inherent polarization modes

As a middle state between conventional vortex beam with integer topological charge and integer-order VVB, vortex beam with fractional topological charge exhibits peculiar propagative behavior in free space. Their polarization states rotate continuously during propagating in free space, while the orders remain. That is, the second scientific problem is that what is the mechanism of polarization rotation of these peculiar vortex beams in free space.

### 2.2.1 Phase evolution of different vortex beams

According to Eq. (2), the order of vortex beam in Eq. (1) is determined by the topological charges of left and right circular polarization modes. Because both modes are special solution of wave function, their vortex phases can maintain stably in free space. That is, one can always obtain a stable *m*+0.5 order of light beam in Eq. (1) due to the unchanged topological charges. Unlike the order of vortex beam in Eq. (1), the polarization rotation of light beam is not dependent on the topological charge, but the phase difference between inherent left and right circular polarization modes. For this reason, the key to reveal the polarization rotation of these peculiar vortex beams is to find out the phase evolution of different vortex beams in Eq. (2).

Theoretically speaking, propagable vortex beams in Eqs. (12, 13) can be considered as two conventional optical vortices with opposite circular polarization and topological cores with a difference of 1. As shown in Fig. 1, points O and $O_1$ are the geometric focus of OL and the points at a bright ring of vortex beam, which can be expressed as $(0,0,0)$, $(\rho_l \cos\varphi_s, \rho_l \sin\varphi_s, z)$, respectively. Here, $z=0$ indicates the focal plane of OL.

As shown in Fig. 1, all light beam at the focal sphere are focused on the geometrical focus O when the topological charge $l$ is zero. Because there is no phase modulation in the wavefront of OL, the phase of light beam is zero in the focal plane. That is, one can obtain the phase of vortex beam with different topological charge by comparing with that of zero topological charge. Specifically, the OPD for the light beams between the point A and O, $O_1$ can be simplified to $L_0$-$L_3$, where $L_0$ and $L_3$ denote the light paths AO and $AO_1$, respectively. As shown in Fig. 1, AO is the radius $f$ of the focal sphere $\Omega$. Generally, $L_0$ and $L_3$ can be expressed as

$$L_0 = f \tag{14}$$

$$L_3 = \sqrt{(f\sin\theta\cos\varphi - \rho_l\cos\varphi_s)^2 + (f\sin\theta\sin\varphi - \rho_l\sin\varphi_s)^2 + (f\cos\theta)^2} \quad . \tag{15}$$

The OPD $L_0$-$L_3$ can therefore be obtained by

$$\Delta\delta = L_0 - L_3 = \frac{-\rho_l^2/f + 2\rho_l \sin\theta\cos(\varphi-\varphi_s)}{1+\sqrt{1+\rho_l^2/f^2 - 2\sin\theta\cos(\varphi-\varphi_s)\rho_l/f}} \tag{16}$$

Because O, $O_1$ are the points on the vortex beam in the vicinity of the focus, $\rho_l \ll f$. That is, $(\rho_l^2)/f \approx 0$, $\rho_l/f \approx 0$. Finally, Equation 16 can be simplified to

$$\Delta\delta = \rho_l \sin\theta\cos(\varphi-\varphi_s). \tag{17}$$

According to the relationship between the phase and OPD, the phase difference between vortex beams with different topological charge can be obtained by $\Delta\psi = k\Delta\delta$, where $k = 2\pi n/\lambda$ is the wavenumber, $n=1$ is the refractive index in the focusing space, see Fig. 1. Using Eq. (10), the phase difference indicated by Eq. (17) can be expressed as

$$\Delta\psi = \pi(M + l/2)\cos(\varphi-\varphi_s). \tag{18}$$

In the case of $l \geq 0$, $\varphi = \varphi_s$. That is,

$$\Delta\psi_{l>0} = \pi(M + l/2). \tag{19}$$

Note that $M=0, 1, 2, 3\ldots$.

In the case of $l<0$, $\varphi = \varphi_s + l\pi\, \varphi_s$. That is,

$$\Delta\psi_{l<0} = \pi(M + l/2)\cos|l|\pi. \tag{20}$$

Note that $M=0, -1, -2, -3\ldots$. Please refer to Eq. (7).

Without loss of generality, we take $M=0$ as example to investigate the phase difference between different vortex beams. Here, $M=0$ indicates the brightest ring in the center, see Fig. 1. Figure 4 presents the phase of vortex beam $\Delta\psi$ with different topological charge at an infinite distance. When the topological charge $l>0$, their phases are $\pi l/2$. However, when the topological charge $l<0$, the phases can be expressed as $\pi l \cos|l|\pi/2$. That is, the phase sign with $l<0$ is changed alternately in Eq. (20). Although the direct relationship between the phases of vortex beams with positive and negative topological charge is not clear, their phases' transmittances are symmetrical about the light beam with $l=0$ in Fig. 4, which can be expressed as

$$T = \exp(i\Delta\psi) = i^{|l|}. \tag{21}$$

That is, there is an additional phase change when vortex beams propagate from the wavefront of OL to an infinite distance in free space, namely the focal plane.

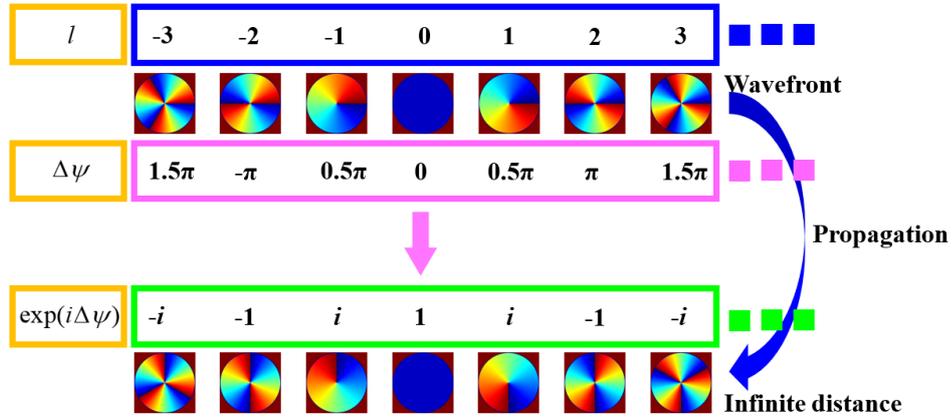

**Figure 4.** Phase evolution of vortex beam with different topological charge in focusing system. Here, the infinite distance denotes the focal plane of OL. $\Delta\psi$ is the changing phase when vortex beam propagates from the wavefront of OL to infinite distance.

### 2.2.2 Phase of vortex beam with different M

Based on the principle of interference, vortex beam with topological charge $l$ possesses not only the brightness ring in center in Fig. 5, but also other bright rings located at the periphery of light beam. Normally, it is not easy to observe the outer rings due to the weak diffraction effect of vortex beam, see Fig. 5(a). However, as shown in Fig. 5 (b, c), when modulating by a high-pass pupil filter or the topological charge is large enough, the diffraction effect of light beam becomes large, and the outer rings cannot be neglected accordingly. To reveal the phase evolution of entire vortex beam, we investigate the phase of different $M$.

Generally, the phase of geometric focus O is zero because there is no phase in the wavefront of OL. Therefore, one can obtain the phase of different $M$ in a vortex beam by comparing with that of zero topological charge as well. Supposed points $O_1$ and $O_2$ in Fig. 5 are two points at two arbitrary bright rings of a vortex beam with topological charge $l$, which can be expressed as $(\rho_{M,l}\cos\varphi_s, \rho_{M,l}\sin\varphi_s, z)$. Here, $M$ denotes the order of bright ring, and $z=0$ indicates the focal plane of OL.

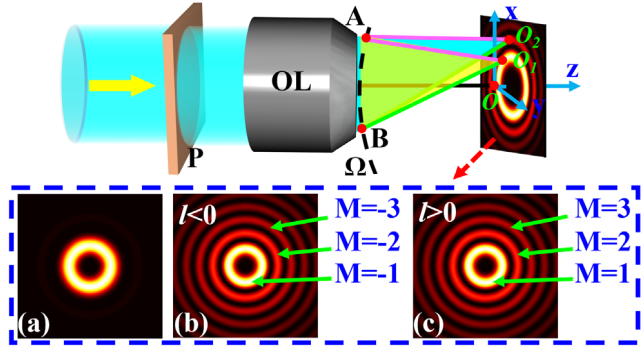

**Figure 5.** Schematic of vortex beam with different $M$ in focusing system. $O_1$ and $O_2$ are two points at two arbitrary bright rings of a vortex beam with topological charge $l$.

Based on a similar process of derivation above, the phase of different $M$ can be obtained by simply calculating the OPD between AO and $AO_{1,2}$, or BO and $BO_{1,2}$. That is, according to Eqs. (18) and (19), the phase of different $M$ can be rewritten as

$$\Delta\phi_{l>0} = \pi(M + l/2) \quad (22)$$

$$\Delta\phi_{l<0} = \pi(M + l/2)\cos|l|\pi \quad (23)$$

, where $\Delta\phi_{l>0}$ and $\Delta\phi_{l<0}$ are the phases of vortex with $l>0$ and $l<0$, respectively. Note that $M$=0, -1, -2, -3…for $l<0$, while $M$=0, 1, 2, 3…for $l>0$, as shown in Fig. 5(b, c).

For a particular $l$, the phase difference between bright ring and focus O can be expressed as

$$\Delta\phi_{l>0} = \pi(M - M_0) \quad (24)$$

$$\Delta\phi_{l<0} = \pi(M - M_0)\cos|l|\pi \quad (25)$$

where $M_0 = 0$ represents the bright focus O in the center of focal plane. As demonstrated in Section 2.2.1, different $l$ denotes different initial phase change with $M$=0. Here, we name this phase as $\phi_o$, which can be found in Fig. 4. Taking $l=-1$ as example. In this case, $\phi_o = 0.5\pi$. Figure 6 presents the phase $\Delta\phi$ of vortex beam with different $M$ at an infinite distance. Although the initial phase is different for different vortex beam, the phase differences between adjacent rings $|\Delta\phi|$ are always equal to π. As shown in Fig. 6, these phase change caused by the propagation of vortex beam does not affect the stability of entire optical vortex, thereby leading to the unchanged topological charge of entire light beam.

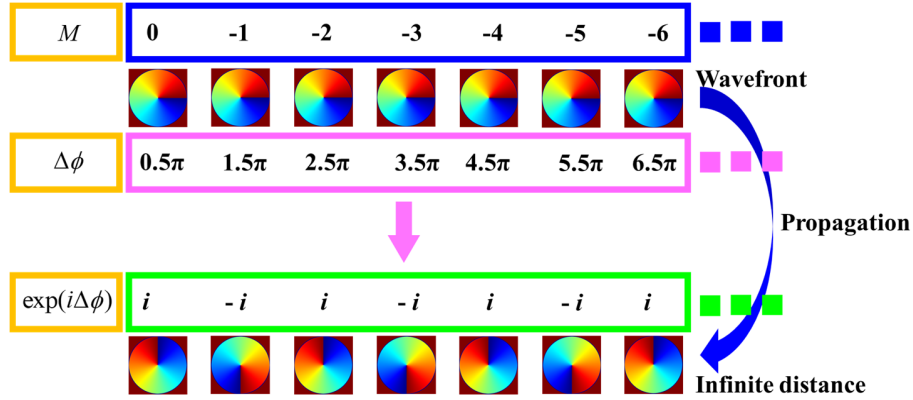

**Figure 6.** Phase evolution of vortex beam with different *M* in focusing system. Here, the infinite distance denotes the focal plane of OL. $\Delta\phi$ is the changing phase when vortex beam propagates from the wavefront of OL to infinite distance.

### 2.2.3 Polarization rotation of vortex beam in Eq. (1)

Sections 2.2.1 and 2.2.2 have demonstrated that the phase of vortex beam is changed continuously during propagating in free space. Therefore, the mathematical form of light beam in the wavefront of OL is slightly different from that of light beam in the focal plane, namely infinite distance. Before explaining the polarization rotation of vortex beam in Eq. (1), we would like to investigate the polarization state of its counterpart with integer order for comparison.

In term of integer-order VVB, its polarization state in the wavefront of OL can be considered as the combination of left and right circular polarization modes with topological charge $\pm m$, which can be expressed as

$$\mathbf{E} = \exp(im\varphi)|\mathbf{R}\rangle + \exp(-im\varphi)|\mathbf{L}\rangle, \tag{26}$$

where $|\mathbf{L}\rangle = \begin{bmatrix} 1 & i \end{bmatrix}^t$ and $|\mathbf{R}\rangle = \begin{bmatrix} 1 & -i \end{bmatrix}^t$ denote the left and right circularly polarized modes, respectively. *t* denotes the matrix transpose operator.

When *m*-order VVB propagates to an infinite distance, a phase change is occurred at the inherent polarization modes $|\mathbf{L}\rangle$ and $|\mathbf{R}\rangle$. According to Eq. (21), its polarization state at infinite distance turns into

$$\mathbf{E} = i^m \exp(im\varphi)|\mathbf{R}\rangle + i^{|-m|}\exp(-im\varphi)|\mathbf{L}\rangle. \tag{27}$$

By comparing with the light beam in Eq. (26), the phase change $i^m = i^{|-m|}$ is independent of polarization state of light beam in Eq. (27). For this reason, the polarization rotation cannot realize in free space, and $m$-order VVB always maintains stably in free space.

In term of propagable vortex beam with fractional topological charge, this peculiar vortex beam has two stable states in free space in Eqs. (12) and (13). Both propagable light beams can also be divided into the combination of left and right circular polarization modes as well. Specifically, VVB with the order of $m+0.5$ can be simplified to

$$\mathbf{E}_{m+0.5,+} = \exp[i(m+1)\varphi]\begin{bmatrix}1\\-i\end{bmatrix} + \exp(-im\varphi)\begin{bmatrix}1\\i\end{bmatrix}, \tag{28}$$

$$\mathbf{E}_{m+0.5,-} = \exp(im\varphi)\begin{bmatrix}1\\-i\end{bmatrix} + \exp[-i(m+1)\varphi]\begin{bmatrix}1\\i\end{bmatrix}, \tag{29}$$

Here, $\mathbf{E}_{m+0.5,+}$ and $\mathbf{E}_{m+0.5,-}$ indicate $m+0.5$-order VVB with topological charge $+0.5$ and $-0.5$, respectively. After propagating to an infinite distance, both $m+0.5$-order VVBs in Eqs. (28) and (29) turn into [see Eq. (23)]

$$\mathbf{E}_{m+0.5,+} = i^{|m+1|}\exp[i(m+1)\varphi]\begin{bmatrix}1\\-i\end{bmatrix} + i^{|-m|}\exp(-im\varphi)\begin{bmatrix}1\\i\end{bmatrix}, \tag{30}$$

$$\mathbf{E}_{m+0.5,-} = i^{|m|}\exp(im\varphi)\begin{bmatrix}1\\-i\end{bmatrix} + i^{|-(m+1)|}\exp[-i(m+1)\varphi]\begin{bmatrix}1\\i\end{bmatrix}. \tag{31}$$

Supposed $m>0$, the above two Equations can be finally simplified to

$$\mathbf{E}_{m+0.5,+} = i^m \exp(i0.25\pi)\exp(i0.5\varphi)\begin{bmatrix}\cos[(m+0.5)\varphi+0.25\pi]\\\sin[(m+0.5)\varphi+0.25\pi]\end{bmatrix}, \tag{32}$$

$$\mathbf{E}_{m+0.5,-} = i^m \exp(i0.25\pi)\exp(-i0.5\varphi)\begin{bmatrix}\cos[(m+0.5)\varphi-0.25\pi]\\\sin[(m+0.5)\varphi-0.25\pi]\end{bmatrix}. \tag{33}$$

Vortex beam with topological charge $l+0.5$ is another stable state of light beam in Eq. (13). Likewise, it can also be expressed as

$$\mathbf{E}_{l+0.5,+} = \exp[i(l+1)\varphi]\begin{bmatrix}1\\-i\end{bmatrix} + \exp(il\varphi)\begin{bmatrix}1\\i\end{bmatrix}, \tag{34}$$

$$\mathbf{E}_{l+0.5,-} = \exp(il\varphi)\begin{bmatrix}1\\-i\end{bmatrix} + \exp[i(l+1)\varphi]\begin{bmatrix}1\\i\end{bmatrix}, \tag{35}$$

That is, when both vortex beams propagate to an infinite distance, their polarization states turn into

$$\mathbf{E}_{l+0.5,+} = i^{|l+1|} \exp[i(l+1)\varphi]\begin{bmatrix}1\\-i\end{bmatrix} + i^{|l|}\exp(il\varphi)\begin{bmatrix}1\\i\end{bmatrix}, \tag{36}$$

$$\mathbf{E}_{l+0.5,-} = i^{|l|}\exp(il\varphi)\begin{bmatrix}1\\-i\end{bmatrix} + i^{|l+1|}\exp[i(l+1)\varphi]\begin{bmatrix}1\\i\end{bmatrix}, \tag{37}$$

Supposed $l>0$, the above two Equations can be finally simplified to

$$\mathbf{E}_{l+0.5,+} = i^{l}\exp[i(l+0.5)\varphi]\exp(i0.25\pi)\begin{bmatrix}\cos(0.5\varphi+0.25\pi)\\ \sin(0.5\varphi+0.25\pi)\end{bmatrix}, \tag{38}$$

$$\mathbf{E}_{l+0.5,-} = i^{l}\exp[i(l+0.5)\varphi]\exp(i0.25\pi)\begin{bmatrix}\cos(-0.5\varphi-0.25\pi)\\ \sin(-0.5\varphi-0.25\pi)\end{bmatrix}. \tag{39}$$

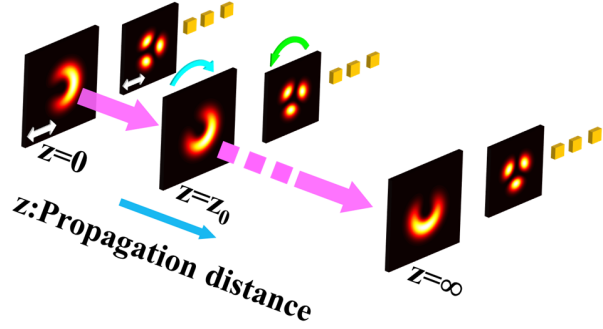

**Figure 7.** Schematic of polarization rotation of vortex beam with fractional topological charge. Here, the blue and green arrow denote the direction of rotation.

According to Equations (32, 33, 38, 39), we present the schematic of polarization rotation in Fig. 7. Owing to the phase change caused by the propagation of vortex beam, the polarization state of $m+0.5$-order VVB are rotating continuously from $\beta=0$ in the wavefront of OL to $\beta=\pm 0.25\pi$ at the focal plane, namely infinite distance. The direction of polarization rotation is determined by the sign of VVB's order $m+0.5$ and $\beta=\pm 0.25\pi$. Specifically, $m\bullet\beta>0$ implies the clockwise rotation of polarization, while $m\bullet\beta<0$ denotes the counterclockwise rotation of polarization, as shown in Fig. 7. Beside the polarization rotation of vortex beam, these peculiar light beams also possess an additional phase change like that of conventional vortex beam, which are changing continuously from 0 in the wavefront of OL to $i^m\exp(i0.25\pi)$ and $i^l\exp(i0.25\pi)$ in Eqs. (32, 33, 38, 39) at the focal plane, namely infinite distance. However, the order and topological charge of light beam remain, thereby making these peculiar vortex beams propagable in free space.

## 2.3 Electric field continuity of fractional-strength optical vortices

As usual, $m+0.5$-order VVB and vortex phase with $l+0.5$ topological charge cannot propagate stable individually in free space. However, when overlapping both unstable states together, one can obtain two propagable vortex beams in Eqs. (12, 13), which possess not only a stable polarization of $m+0.5$-order VVB, but also a stable vortex phase with $l+0.5$ topological charge. Why can the interaction between the two unstable states maintain the entire propagable vortex beams in Eqs. (12, 13) stably in free space? Why is the discontinuity of light beam caused by the phase and polarization jump missing? In this section, we would like to solve this third scientific problem by revealing the relationship between the phase domain and polarization domain.

### 2.3.1 Phase's binary time vector property

In principle, a light beam propagating along $+z$ axis can be expressed in a generalized form:

$$\mathbf{E} = A\cos(\phi + \omega t)\mathbf{E}_0, \tag{40}$$

where A is the amplitude of light beam; $\phi = kz$ denotes the phase of entire light beam, $k = 2\pi n/\lambda$ is the wavenumber, $n=1$ is the refractive index in free space. $\omega = 2\pi/T$, $T$ is the period of electric field of light beam. $\mathbf{E}_0$ denote the polarization state of light beam. For example, one can obtain a conventional vortex beam with $\phi = l\varphi$ and $\mathbf{E}_0 = \begin{bmatrix} 1 & 0 \end{bmatrix}^t$. $\mathbf{E}_0 = \begin{bmatrix} 1 & 0 \end{bmatrix}^t$ indicates a $x$ linearly polarized beam.

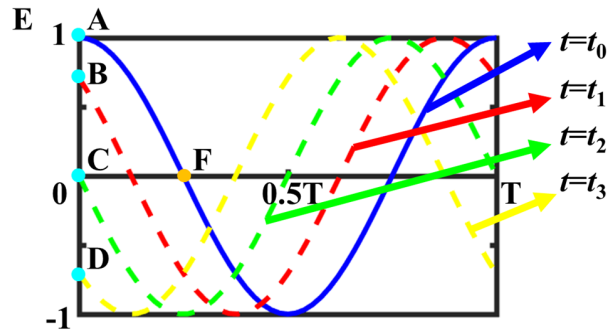

**Figure 8.** Schematic of electric field oscillation over time $t$ for a single light beam. T indicates the period of the entire oscillation.

As show in Fig. 8, Equation 40 represents the oscillation of electric field over time $t$ for a single light beam when $z = z_0$. The adjustment of the phase $\phi$ does not affect the waveform of oscillation, but the starting point of electric field at $t$, see points A, B, C, D in Fig. 8. That is, one can obtain the starting point of electric field at $t_2$ (Point C) by controlling the phase $\phi = kz$ with a corresponding optical path $z = 0.25\lambda$ at $t_0$ (Point F). For this reason, the phase of light beam actually represents the starting point of electric field oscillation of light beam at a particular $t$. From this viewpoint, it is not reasonable to consider the phase alone by ignoring the entire electric field in Eq. (40). In other word, one must consider $\exp(il\varphi)\mathbf{E}_0$ as a whole for a light beam instead of merely the phase $\exp(il\varphi)$.

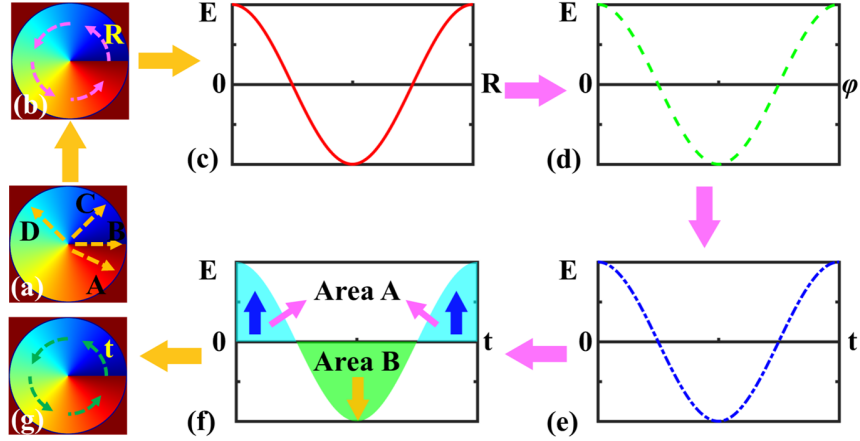

**Figure 9.** Schematic of phase's binary time vector property.

In term of vortex beam with topological charge $l$, the entire light beam can be considered as the combination of many different light beams modulated by the vortex phase $\phi = l\varphi$, see Fig. 9 (a). Taking $l$=1 as example in Fig. 9. Owing to the spatial variant of phase along the radius R in Fig. 9 (b), different starting point of electric field oscillation that change continuously can be obtained in the transverse plane of vortex beam simultaneously in Fig. 9 (c, d). Note that the period in Fig. 9 (c, d) is merely determined by the topological charge $l$. For more clear, one can turn electric field oscillation in Fig. 9 (d) into time domain. Here, the x-coordinate $t$ in Fig. 9(e, f) is corresponding to different starting point of electric field oscillation like that in Fig. 8. According to Eq. (40), the phase represents the electric field oscillation of light beam at a particular $t$. For an entire oscillation period $T$, the electric field of light beam can be divided into two areas along the radius R in Fig. 9 (f), where the directions of electric fields are upward indicated by blue arrow in area A and downward indicated by yellow arrow in area B, respectively. Both inverse electric fields demonstrate that the oscillation of electric field caused by the change of vortex phase can lead to the reversal of polarization. It should be emphasized that the vector property induce by the change of phase merely possesses a binary direction, as shown in Fig. 9 (f). Here, we call this property of light beam as phase's binary time vector property.

**2.3.2 Coexistence of phase and polarization discontinuity**

According to Eq. (40), $\mathbf{E}_0$ denote the polarization state of light beam. For example, $\mathbf{E}_0 = \begin{bmatrix} 1 & 0 \end{bmatrix}^t$ denotes linear polarization state of light beam, while $\mathbf{E}_0 = \begin{bmatrix} \cos m\varphi & \sin m\varphi \end{bmatrix}^t$ indicates the polarization state of $m$-order VVB. From both examples, one can find that the difference between the phase and polarization state of light beam. Specifically, as shown in Fig. 10, the phase represents the electric field oscillation of a light beam indicated by the red arrow, while its oscillation trajectory is determined by the unit vector $\mathbf{E}_0$, namely the direction of electrics field indicated by the blue arrow. In principle, electric field oscillation is time-variant that the electric field of light beam is changing periodically with time $t$ in Fig. 9, while the polarization state is spatial-variant. For a stable polarized beam, the electric field oscillation in its transverse plane are synchronous not matter what the spatial distribution of polarization state is, such as $m$-order VVB.

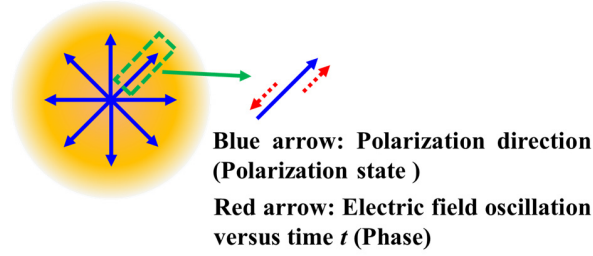

**Figure 10.** Schematic of polarization and phase function of light beam.

Normally, when modulating the phase of $m$-order VVB, the polarization of original $m$-order VVB cannot maintain because the phase affects the entire electric field in Eq. (40). However, in term of propagable vortex beams in Eqs. (12, 13), the stability of vortex beam is dependent on the interaction between the vortex phase with topological charge $l+0.5$ and polarization with $m+0.5$-order VVB. As shown in Fig. 11 (a, b), there are polarization discontinuity and phase jump along x axis. For the polarization of $m+0.5$-order VVB in Fig. 11 (a), the electric field oscillation along x axis has two inverse oscillation processes. As shown in Fig. 11 (b), one is indicated by the red curve; another is the blue curve. Because both processes are synchronous [see blue and yellow arrow in Fig. 11 (d)], a polarization discontinuity is always obtained along x axis, and the light intensity along x axis is zero accordingly. For the vortex phase with topological charge $l+0.5$ in Fig. 11 (b), owing to the property of Phase's binary time vector property, the oscillation processes along x axis are formed by the combination of two inverse electric fields. For the electric field oscillation indicated by the green arrow in Fig. 11(b), its starting point are located at the area A in Fig. 11(e). That is, at $t=0$, the direction of electric field is upward indicated by blue arrow. Likewise, the starting point of its inverse counterpart indicated by the green arrow is corresponding to that of blue arrow at time $t=0$ with $\phi=\pi$. Therefore, the direction of electric field is downward. Because of both inverse electric field caused by the property of Phase's binary time vector property, a dark line along x axis is created in the transverse plane of light beam.

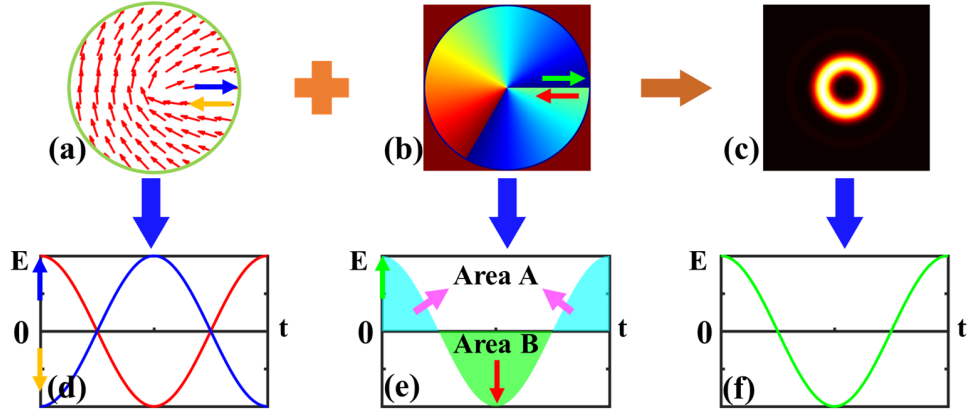

**Figure 11.** Schematic of coexistence of phase and polarization discontinuity.

The polarization of *m*+0.5 order VVB in Fig. 11 (a) and the vortex phase with topological charge *l*+0.5 in Fig. 11 (b) can induce polarization discontinuity and phase jump along x axis, respectively. However, when overlapping together, the upward starting point of electric field in Fig. 11 (d) (blue arrow) and in Fig. 11 (e) (green arrow) remain unchanged as before. On the contrary, their downward counterparts (see yellow arrow in Fig. 11 (d) and red arrow in Fig. 11 (e)) are interacted with each other. That is, the combined starting point are reversed, thereby leading to an identical electric field oscillation process in Fig. 11 (f). For the above reasons, the dark line along x axis caused by both discontinuities is disappeared in Fig. 11 (c), and the electric field is changing continuously in the transverse plane of propagable vortex beam in Eq. (1). Moreover, the polarization and phase are in the framework of spatial and time domain, respectively. That is why the phase and polarization discontinuity in Fig. 11 (a, b) can coexist within the propagable vortex beam in Eq. (1).

## 3. Conclusion

In conclusion, we have revealed the physical essence of propagable vortex beam with *l*+0.5 topological charge by re-understanding Abbe's diffraction limit, investigating the phase evolution of vortex beam and proposing phase's binary time vector property. For the first one, because of Abbe's diffraction limit, the fractional-order VVBs can still remain stable even though their modulation symmetries are naturally broken down by the optical vortices with fractional topological charge. For the second one, we demonstrate the phase evolution of vortex beam, which is the physical reason of polarization rotation of fractional-order VVBs. For the third one, the phase is not merely a scalar attribute of light beam, but manifests a binary time vector property. Therefore, the electric field of entire fractional-order VVBs is continuous while the topological charge of vortex phase and the order of polarization are fraction. This work provides entirely different physical viewpoints on the phase of light beam and Abbe's diffraction limit, which may deepen our knowledge on the behavior of light beam in classical optics.

## Data Availability

All data supporting the findings of this study are available from the corresponding author on request.

**References**

1. L. Allen, M. W. Beijersbergen, R. J. C. Spreeuw, J. P. Woerdman, Orbital angular momentum of light and the transformation of Laguerre-Gaussian laser modes. *Phys. Rev. A* **45**, 8185-8189 (1992).
2. Q. Zhan, Cylindrical vector beams: from mathematical concepts to applications. *Adv. Opt. Photon.* **1**, 1-57 (2009).
3. J. Wang, J. Yang, I. M. Fazal, N. Ahmed, Y. Yan, H. Huang, Y. Ren, Y. Yue, S. Dolinar, M. Tur and A. E. Willner, Terabit free-space data transmission employing orbital angular momentum multiplexing. *Nat. Photonics* **6**, 488-496 (2012).
4. R. C. Devlin, A. Ambrosio, N. A. Rubin, J. P. B. Mueller, F. Capasso, Arbitrary spin-to–orbital angular momentum conversion of light. *Science* **358**, 896-901 (2017).
5. T. Stav, A. Faerman, E. Maguid, D. Oren, V. Kleiner, E. Hasman, M. Segev, Quantum entanglement of the spin and orbital angular momentum of photons using metamaterials. *Science* **361**, 1101-1103 (2018).
6. J. Courtial, D. A. Robertson, K. Dholakia, L. Allen, M. J. Padgett, Rotational Frequency Shift of a Light Beam. *Phys. Rev. Lett.* **81**, 4828-4830 (1998).
7. G. Li, T. Zentgraf, S. Zhang, Rotational Doppler effect in nonlinear optics. *Nat. Phys.* **12**, 736-740 (2016).
8. D. Naidoo, F. S. Roux, A. Dudley, I. Litvin, B. Piccirillo, L. Marrucci, A. Forbes, Controlled generation of higher-order Poincaré sphere beams from a laser. *Nat. Photonics* **10**, 327-332 (2016).
9. P. Gao, B. Prunsche, L. Zhou, K. Nienhaus, G. U. Nienhaus, Background suppression in fluorescence nanoscopy with stimulated emission double depletion. *Nat. Photonics* **11**, 163-169 (2017).
10. N. Bozinovic, Y. Yue, Y. Ren, M. Tur, P. Kristensen, H. Huang, A. E. Willner, S. Ramachandran, Terabit-Scale Orbital Angular Momentum Mode Division Multiplexing in Fibers. *Science* **340**, 1545-1548 (2013).
11. X. Wang, X. Cai, Z. Su, M. Chen, D. Wu, L. Li, N. Liu, C. Lu and J. Pan, Quantum teleportation of multiple degrees of freedom of a single photon. *Nature* **518**, 516-519 (2015).
12. X. Weng, Y. Miao, G. Wang, Q. Zhan, X. Dong, J. Qu, X. Gao, S. Zhuang, Light beam carrying natural non-integer orbital angular momentum in free space. arXiv:2105.11251 [physics.optics].
13. Y. Miao, X. Weng, Y. Wang, L. Wang, G. Wang, X. Gao, S. Zhuang, Parallel creation of propagable integer- and fractional-order vector vortex beams using mode extraction principle. *Opt. Lett.* **47**, 3319-3322 (2022).
14. X. Weng, Y. Miao, Q. Zhang, G. Wang, Y. Li, X. Gao, S. Zhuang, Extraction of Inherent Polarization Modes from an m-Order Vector Vortex Beam. *Adv. Photonics Res.* 2100194, 2022.
15. 苏显渝，吕乃光，陈家璧，《信息光学原理》. Chapter 3.
16. G. Milione, M. P. J. Lavery, H. Huang, Y. Ren, G. Xie, T. A. Nguyen, E. Karimi, L. Marrucci, D. A. Nolan, R. R. Alfano, A. E. Willner, 4 x 20Gbit/s mode division multiplexing over free space using vector modes and a q-plate mode (de)multiplexer. *Opt. Lett.* **40**, 1980-1983 (2015).

**Acknowledgments**

Parts of this work were supported by the National Natural Science Foundation of China (62022059/11804232) and the National Key Research and Development Program of China (2018YFC1313803).

**Author contributions**

X. Weng conceived of the research and carried out the numerical simulations. X. Weng and X. Gao cowrote the paper with support from Y. Miao, X. Dong, Y. Li. X. Gao and S. Zhuang offered advice regarding development. S. Zhuang directed the entire project. All of the authors participated in the analysis and discussion of the results.

**Competing interests statement**

The authors declare that they have no competing financial or nonfinancial interests to disclose.